\documentclass[twocolumn,superscriptaddress,showkeys]{revtex4}
\usepackage{amssymb,epsf,epsfig,amsmath,color, array}

\newcommand{\lsim}{
\mathrel{\hbox{\rlap{\hbox{\lower4pt\hbox{$\sim$}}}\hbox{$<$}}}}
\newcommand{\gsim}{
\mathrel{\hbox{\rlap{\hbox{\lower4pt\hbox{$\sim$}}}\hbox{$>$}}}}

\def\D0{D\O }

\begin{document}
\begin{titlepage}
\vspace*{1.7truecm}
\begin{flushright}
Nikhef-2010-048
\end{flushright}

\vspace{1.6truecm}

\begin{center}
\boldmath
{\Large{\bf Tests of Factorization and $SU(3)$ Relations

\vspace*{0.3truecm}

in $B$ Decays into Heavy-Light Final States}}

\unboldmath
\end{center}

\vspace{1.2truecm}

\begin{center}
{\bf Robert Fleischer, Nicola Serra and Niels Tuning}

\vspace{0.5truecm}

{\sl Nikhef, Science Park 105, NL-1098 XG Amsterdam, The Netherlands}

\end{center}

\vspace*{1.7cm}

\begin{center}
\large{\bf Abstract}\\

\vspace*{0.6truecm}

\begin{tabular}{p{14.5truecm}}
Using data from the $B$ factories and the Tevatron, we perform tests of how well 
nonleptonic $B$ decays of the kind $B\to D^{(*)}_{(s)} P$, where $P$ is a 
pion or kaon, are described within the factorization framework. We find that factorization 
works well -- as is theoretically expected -- for color-allowed, tree-diagram-like topologies.
Moreover, also exchange topologies, which have a nonfactorizable character, do 
not show any anomalous behavior. We discuss also isospin triangles between the 
$B\to D^{(*)}\pi$ decay amplitudes, and determine the corresponding amplitudes 
in the complex plane, which show a significant enhancement of the color-suppressed
tree contribution with respect to the factorization picture. Using data for $B\to D^{(*)}K$
decays, we determine $SU(3)$-breaking effects and cannot resolve any nonfactorizable
$SU(3)$-breaking corrections larger than $\sim5\%$. In view of these results, we point 
out that a comparison between the $\bar B^0_d\to D^+\pi^-$ and $\bar B^0_s\to D_s^+\pi^-$ 
decays offers an interesting new determination of $f_d/f_s$. Using CDF data, we obtain the most
precise value of this ratio at CDF, and discuss the prospects for a corresponding  
measurement at LHCb.
\end{tabular}

\end{center}

\vspace*{3.7truecm}

\vfill

\noindent
December 2010

\end{titlepage}

\newpage
\thispagestyle{empty}
\mbox{}

\newpage
\thispagestyle{empty}
\mbox{}

\rule{0cm}{23cm}

\newpage
\thispagestyle{empty}
\mbox{}

\setcounter{page}{0}

\preprint{Nikhef-2010-048}

\date{\today}

\title{\boldmath Tests of Factorization and $SU(3)$ Relations
in $B$ Decays into Heavy-Light Final States\unboldmath}

\author{Robert Fleischer}
\affiliation{Nikhef, Science Park 105, NL-1098 XG Amsterdam, The Netherlands}

\author{Nicola Serra}
\affiliation{Nikhef, Science Park 105, NL-1098 XG Amsterdam, The Netherlands}

\author{Niels Tuning}
\affiliation{Nikhef, Science Park 105, NL-1098 XG Amsterdam, The Netherlands}

\begin{abstract}
\vspace{0.2cm}\noindent
Using data from the $B$ factories and the Tevatron, we perform tests of how well 
nonleptonic $B$ decays of the kind $B\to D^{(*)}_{(s)} P$, where $P$ is a 
pion or kaon, are described within the factorization framework. We find that factorization 
works well -- as is theoretically expected -- for color-allowed, tree-diagram-like topologies.
Moreover, also exchange topologies, which have a nonfactorizable character, do 
not show any anomalous behavior. We discuss also isospin triangles between the 
$B\to D^{(*)}\pi$ decay amplitudes, and determine the corresponding amplitudes 
in the complex plane, which show a significant enhancement of the color-suppressed
tree contribution with respect to the factorization picture. Using data for $B\to D^{(*)}K$
decays, we determine $SU(3)$-breaking effects and cannot resolve any nonfactorizable
$SU(3)$-breaking corrections larger than $\sim5\%$. In view of these results, we point 
out that a comparison between the $\bar B^0_d\to D^+\pi^-$ and $\bar B^0_s\to D_s^+\pi^-$ 
decays offers an interesting new determination of $f_d/f_s$. Using CDF data, we obtain the most
precise value of this ratio at CDF, and discuss the prospects for a corresponding  
measurement at LHCb.
\end{abstract}

\maketitle

\section{Introduction}
\begin{figure}[!b]
  \begin{center}
    \begin{picture}(250,80)(0,0)
      \put(  0,0){\includegraphics[scale=0.15]{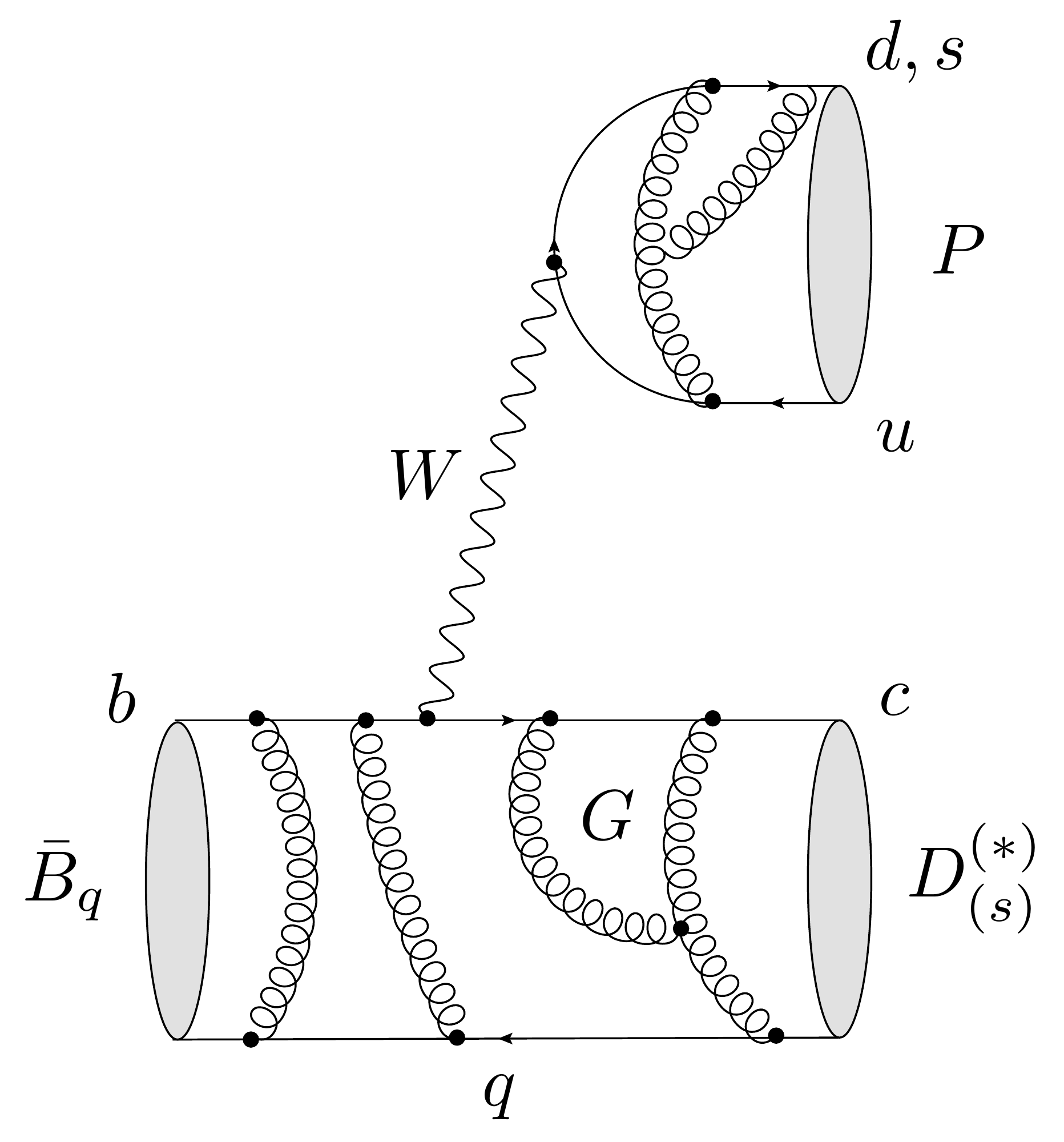}}
      \put( 20,80){T}
      \put( 79,0){\includegraphics[scale=0.15]{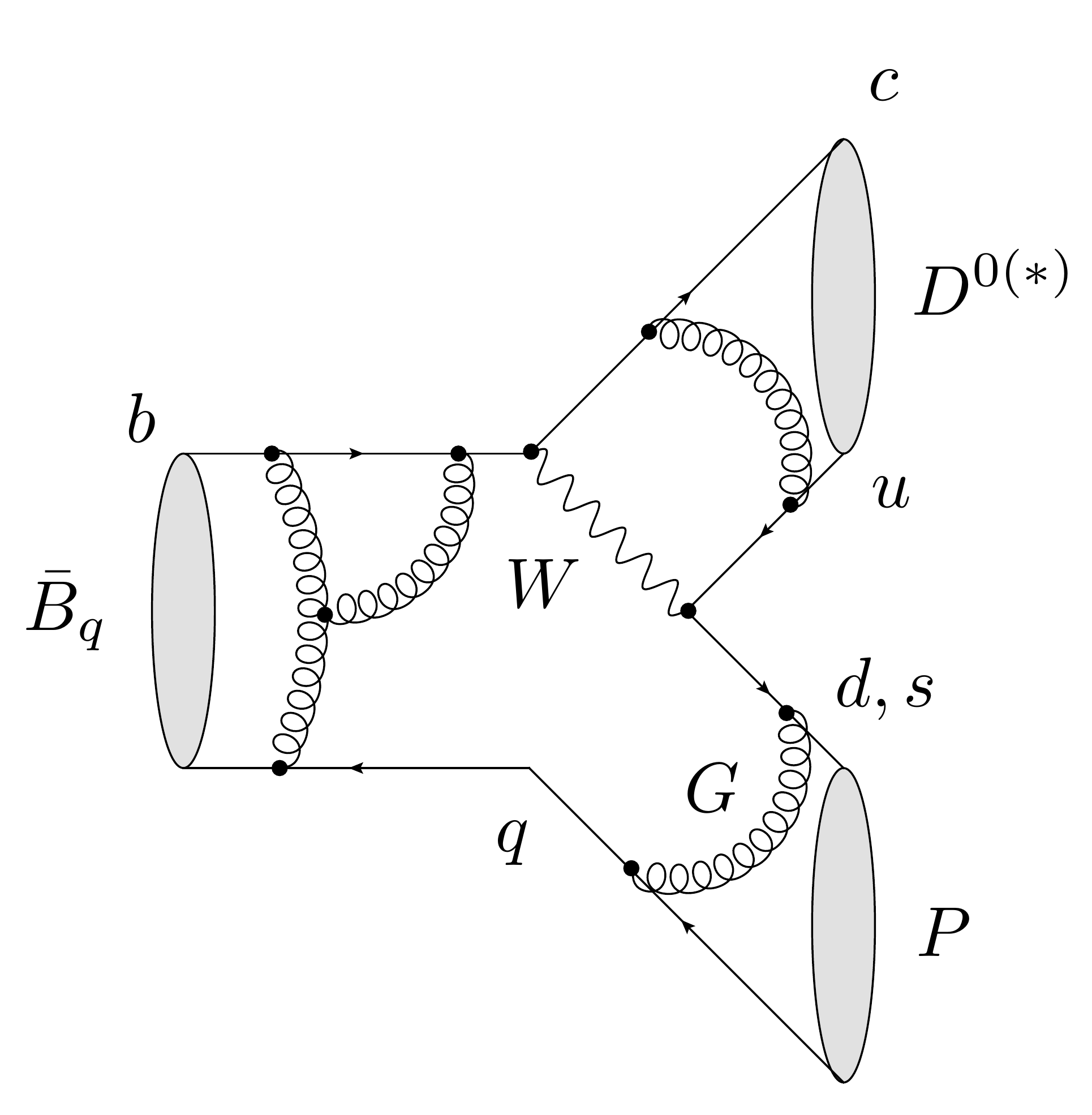}}
      \put( 99,80){C}
      \put(161,0){\includegraphics[scale=0.15]{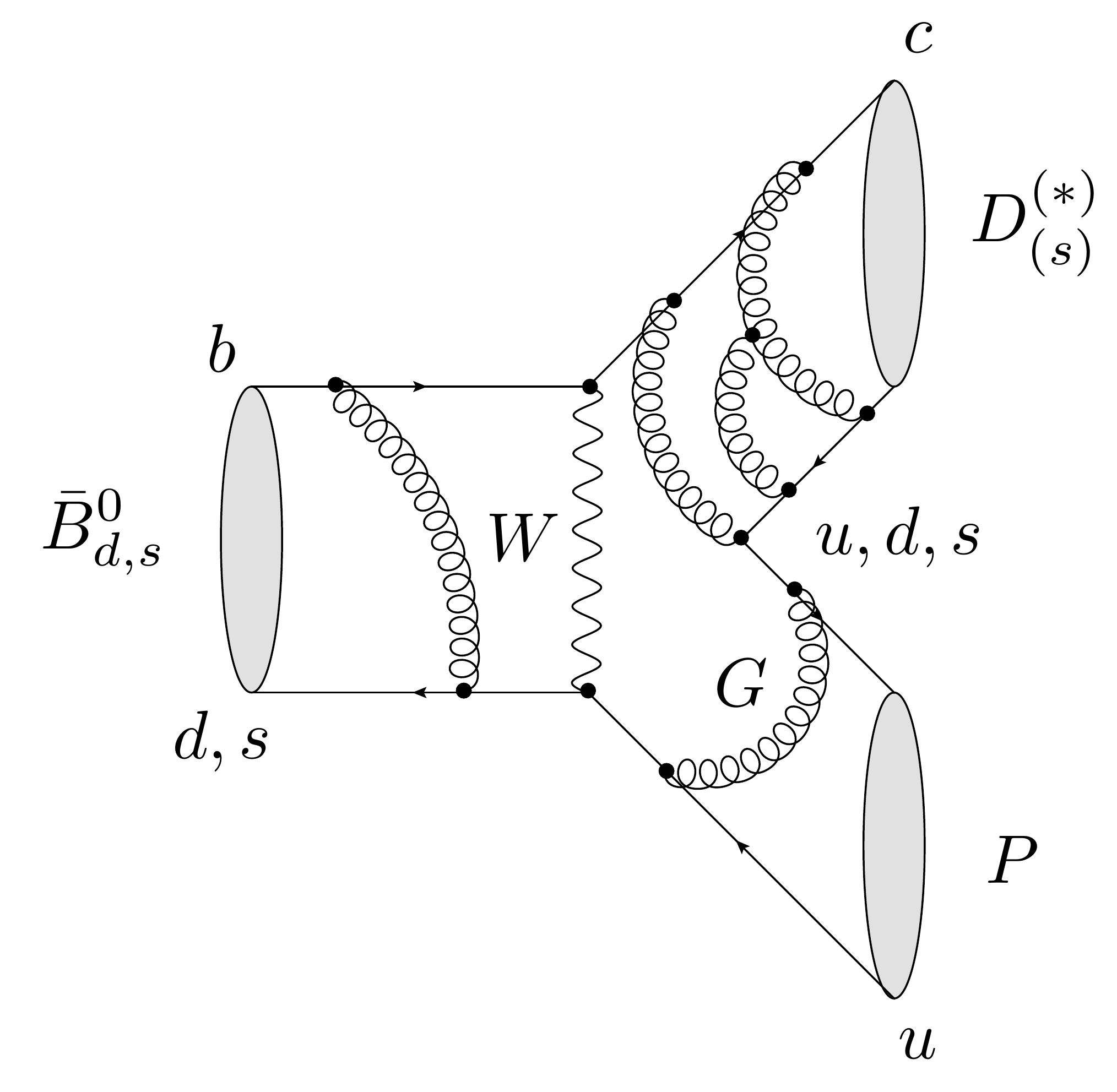}}
      \put(181,80){E}
    \end{picture}
    \caption[TCE]{\em The color-allowed (tree) ($T$),  color-suppressed ($C$) and 
exchange ($E$) topologies contributing to heavy-light decays ($q\in\{u,d,s\}$).}
    \label{fig:top}
  \end{center}
\end{figure}
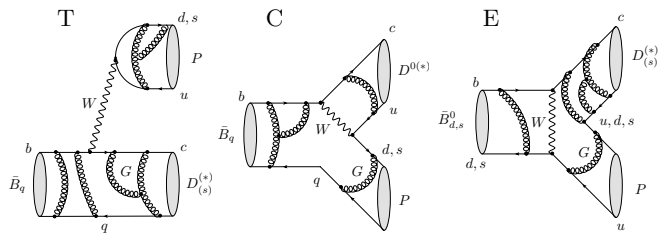

Nonleptonic weak decays of $B$ mesons play an outstanding role for the exploration
of flavor physics and strong interactions. The key challenge of their theoretical description  
is related to the fact that the corresponding low-energy effective Hamiltonian contains
local four-quark operators. Consequently, in the calculation of the transition amplitude, 
we have to deal with nonperturbative ``hadronic" matrix elements of these operators.
For decades we have applied the ``factorization'' hypothesis, i.e.\ to estimate
the matrix element of the four-quark operators through the product of the matrix elements
of the corresponding quark currents \cite{fact}. In the 1980s, the $1/N_{\rm C}$-expansion 
of QCD \cite{BGR} and ``color transparency" arguments \cite{bjor,DG} were used to 
justify this concept, while it could be put on a rigorous theoretical basis in the heavy-quark 
limit for a variety of $B$ decays about ten years ago \cite{BBNS,SCET}. A very useful 
approach to deal with nonleptonic decays is provided by the decomposition of their
amplitudes in terms of different decay topologies  and to apply the $SU(3)$ flavor symmetry
of strong interactions to derive relations between them \cite{GHLR}. We shall use the same 
notation as introduced in that paper to distinguish between color-allowed ($T$), 
color-suppressed ($C$) and exchange ($E$) topologies, which are shown in Fig.~\ref{fig:top}.
For a detailed discussion of the connection between this diagrammatic approach and
the low-energy effective Hamiltonian description, the reader is referred to Ref.~\cite{BS}.

Factorization is not a universal feature of nonleptonic $B$ decays and there are 
cases where it is not expected to work. In fact, nonfactorizable effects are also required
to cancel the renormalization-scale dependence in the calculation of the
transition amplitude by means of the low-energy effective Hamiltonian. 
The $B$-factory data also have shown that nonfactorizable effects can indeed play a 
significant role, in particular for large CP-conserving strong phases and direct CP violation. 
In the framework developed in Refs.~\cite{BBNS,SCET}, such effects are described 
by $\Lambda_{\rm QCD}/m_b$ corrections, which are nonperturbative quantities and
can therefore only be estimated theoretically with large uncertainties.

Prime examples where factorization is expected to work well are given by the 
decays $\bar B^0_d\to D^{(*)+}K^-$, which receive only contributions from color-allowed 
tree-diagram-like topologies. In Ref.~\cite{FST}, we have exploited this feature to propose
a new strategy to determine the ratio $f_d/f_s$ of the fragmentation functions, which
describe the probability that a $b$ quark will fragment in a $\bar B_{d,s}$ meson. It uses
the decays $\bar B^0_d\to D^{+}K^-$ and $\bar B^0_s\to D_s^+\pi^-$. Since the ultimate
precision is limited by nonfactorizable $U$-spin-breaking corrections, which are 
theoretically expected at the few-percent level in these decays, it is interesting to get
experimental insights into factorization and $SU(3)$-breaking corrections. The ratio 
$f_d/f_s$ enters the measurement of any $B_s$ branching ratio at LHCb and is -- in 
particular -- the major limiting factor for the search of New-Physics signals through 
$\mbox{BR}(B^0_s\to \mu^+\mu^-)$.

In this paper, we would like to use the currently available $B$-factory data to check how 
well factorization works. Factorization tests in $B$ decays into heavy-light final states have 
been studied before, but the precision of the corresponding input data
has now reached a level to obtain a significantly sharper picture. 

The outline is as follows: in Section~\ref{sec:T}, we discuss factorization tests for the 
color-allowed amplitude $T$. In Section~\ref{sec:E}, we constrain the impact of exchange 
topologies, $E$, which do {\it not} factorize, and determine their relative orientation with 
respect to $T$. In Section~\ref{sec:iso}, we use an isospin triangle construction to determine 
also the color-allowed amplitude $C$, while we focus on tests of the $SU(3)$ flavor
symmetry in Section~\ref{sec:su3}. In Section~\ref{sec:appl}, we propose an application 
of these studies, which is a determination of $f_d/f_s$ by means of the ratio of the branching 
ratios of the $\bar B^0_d\to D^{+}\pi^-$ and $\bar B^0_s\to D_s^{+}\pi^-$ decays, and
discuss the implications of CDF data and the prospects for the corresponding measurement
at LHCb. Finally, we summarize our conclusions in Section~\ref{sec:concl}. The input
parameters for our numerical analysis are collected in Table~\ref{tab:input-par}.

\begin{table}[!t]
\begin{center}
\begin{tabular}{|lr|lr|} 
\hline
            &              &           &           \\
\hline
$m_{B^0}$   & 5279.17 MeV  &  $m_{B^0_s}$               & 5336.3  MeV \\
$m_{D^+}$   & 1869.60 MeV  &  $m_{D^+_s}$               & 1968.47 MeV \\
$m_{D^0}$   & 1864.83 MeV  &  $m_{D^{*0}}$              & 2006.96 MeV \\
$m_{D^{*+}}$& 2010.25 MeV  &  $m_{D^{*+}_s}$            & 2112.3  MeV \\
$m_{K^+}$   & 497.61  MeV  &  $m_{K^0}$                 & 493.68  MeV \\
$m_{\pi^+}$ & 139.57  MeV  &  $m_{\pi^0}$               & 134.98  MeV \\
$f_\pi$     & 130.41  MeV  &  $f_K$                     & 156.1   MeV \\
$f_\rho$    & 215     MeV  &  $f_D$                     & 206.7   MeV \\
$f_{D^*}$   & 206.7   MeV  &  $f_{D_s}$                 & 257.5   MeV \\
$\tau_{B^0}$& 1.525 ps     &  $\tau_{B^\pm}/\tau_{B^0}$ & 1.071       \\
$|V_{ud}|$  & 0.97425      &  $|V_{us}|$                & 0.2252      \\
\hline
\end{tabular}
\caption[Parameters]
{\em Parameters used in the numerical analysis.}
\label{tab:input-par}
\end{center}
\end{table}

\boldmath
\section{Information on $T$}\label{sec:T}
\unboldmath
Let us start our discussion by having a closer look at the decays
$\bar B^0_d\to D^{(*)+}K^-$, which receive only contributions
from color-allowed tree-diagram-like topologies $T^{(*)}$.  The Particle Data Group (PDG)
gives the branching ratios $\mbox{BR}(\bar B^0_d\to D^{+}K^-)=(2.0\pm0.6)\times10^{-6}$ 
and $\mbox{BR}(\bar B^0_d\to D^{*+}K^-)=(2.14\pm0.16)\times10^{-4}$ \cite{PDG}.
Using the differential rates of semi-leptonic decays, we can actually probe nonfactorizable 
terms \cite{bjor}. The corresponding expression can be written as follows \cite{BBNS}:
\begin{eqnarray}
\lefteqn{R_P^{(*)}\equiv\frac{\mbox{BR}(\bar B^0_d \to D^{(*)+}P^-)
}{d\Gamma(\bar B^0_d\to D^{(*)+}\ell^-\bar\nu_\ell)
/dq^2|_{q^2=m_P^2}}}\nonumber\\
&&=6\pi^2\tau_{B_d}|V_{P}|^2f_P^2|a_1(D_qP)|^2X_P,
\label{eq:a1} 
\end{eqnarray}
where $\tau_{B_d}$ is the $B_d$ lifetime, $q^2$ the four-momentum transfer to the lepton-pair,
$|V_{P}|$ the corresponding element of the
Cabibbo--Kobayashi--Maskawa (CKM) matrix, $f_P$ is the decay constant of the $P$ 
meson, and $X_P$ deviates from 1 below the percent level. The quantity $a_1(D_qP)$ 
describes the deviation from naive factorization. As discussed in detail in Ref.~\cite{BBNS}, 
this parameter is found in ``QCD factorization" 
as a quasi-universal quantity $|a_1|\simeq 1.05$ with very small 
process-dependent  ``nonfactorizable" corrections. 

A first implementation of the factorization test in (\ref{eq:a1}) for the
$\bar B^0_d\to D^{*+}\pi^-$ channel was performed in Ref.~\cite{BStone}. In the 
last decade, we have seen a lot of progress with the measurements of the
semi-leptonic $\bar B^0_d\to D^{(*)+}\ell^-\bar\nu_\ell$ decays, which play a key role
for the determination of $|V_{cb}|$, and of the nonleptonic $B\to D_{(s)}^{(*)}P$ decays.
The averages of the total exclusive semi-leptonic branching fractions amount to
BR$(\bar B\to D\ell^-\bar\nu_\ell)$=$(2.17 \pm 0.12)\%$
and
BR$(\bar B^0_d\to D^{*+}\ell^-\bar\nu_\ell)$=$(5.05\pm 0.12)\%$~\cite{PDG}.

To parametrize the form factors, usually the variable 
\begin{equation}
w\equiv v\cdot v'=\frac{m_B^2+m_D^2 - q^2}{2m_B m_D}
\end{equation}
is used, which is the product of the four-velocities $v$ and $v'$ of the $B$ and
$D_{(s)}^{(*)}$ mesons, respectively. The correspondence between the
differential rates is given by 
\begin{equation}
\frac{d\Gamma}{dq^2} = \frac{1}{2m_Bm_D} \frac{d\Gamma}{dw}.
\end{equation}

\begin{table}[!t]
\begin{center}
\begin{tabular}{|l|l l|l l|l l|} 
\hline
$\bar B^0_d\to D^{+}\ell^-\bar\nu_\ell$  & \multicolumn{2}{c|}{HFAG~\cite{bib:HFAG}} 
& \multicolumn{2}{c|}{Belle~\cite{bib:HFAG,SemilepBelleD}} 
& \multicolumn{2}{c|}{BaBar~\cite{SemilepBaBarD}} \\
\hline
 $F(1) |V_{cb}|$[10$^{-3}$]& 42.3   & $\pm$ 1.48  & 40.85 & $\pm$7.0 & 43.0 & $\pm$2.15   \\ 
 $\rho^2$                  & 1.18   & $\pm$ 0.06  & 1.12  & $\pm$0.26& 1.20 & $\pm$0.10   \\ 
\hline
\hline
$\bar B^0_d\to D^{*+}\ell^-\bar\nu_\ell$   & \multicolumn{2}{c|}{HFAG~\cite{bib:HFAG}} 
& \multicolumn{2}{c|}{Belle~\cite{SemilepBelleDst}} 
& \multicolumn{2}{c|}{BaBar~\cite{SemilepBaBarDst}}\\
\hline
  $F(1) |V_{cb}|$[10$^{-3}$]& 36.04 & $\pm$0.52   & 34.6  & $\pm$1.0   & 34.4  & $\pm$1.2     \\ 
  $\rho^2$                  & 1.24  & $\pm$0.04   & 1.214 & $\pm$0.034 & 1.191 & $\pm$0.056   \\ 
  $R_1$                     & 1.410 & $\pm$0.049  & 1.401 & $\pm$0.034 & 1.429 & $\pm$0.075   \\ 
  $R_2$                     & 0.844 & $\pm$0.027  & 0.864 & $\pm$0.024 & 0.827 & $\pm$0.043   \\ 
\hline
\end{tabular}
\caption[Caprini parameters]
{\em The parameters for the form-factor parametrization of  Ref.~\cite{Caprini},
as determined by the Belle and BaBar collaborations, and the world average given by 
the Heavy Flavour Averaging Group (HFAG).
The recent precise determination of the form-factor parametrization for 
$\bar B^0_d\to D^{*+}\ell^-\bar\nu_\ell$ presented by the Belle collaboration \cite{SemilepBelleDst} 
is not taken into account in the world average yet.
}\label{tab:par}
\end{center}
\end{table}

\begin{table}[!b]
\begin{center}
\begin{tabular}{|l||l|c|c|} 
\hline
Corresponding               &        & \multicolumn{2}{c|}{$d\Gamma/dq^2 (\times 10^3)$[GeV$^{-2}$ps$^{-1}$]} \\ \cline{2-4}
hadronic decay              & $w$    & BaBar      & Belle              \\ 
\hline
$\bar B^0_d\to D^{+}\pi^-$  & 1.588  & 2.34$\pm$0.13  & 2.36$\pm$0.42  \\ 
$\bar B^0_d\to D^{+} K^- $  & 1.577  & 2.31$\pm$0.13  & 2.32$\pm$0.42  \\ 
\hline				       						    
$\bar B^0_d\to D^{*+}\pi^-$ & 1.503  & 1.99$\pm$0.13  & 1.86$\pm$ 0.09 \\ 
$\bar B^0_d\to D^{*+} K^- $ & 1.492  & 2.08$\pm$0.14  & 1.95$\pm$ 0.10 \\ 
\hline
\end{tabular}
\caption[dG/dq2]
{\em The semi-leptonic differential  decay rates at the values of 
the relevant four-momentum transfers entering the factorization test in Eq.~(\ref{eq:a1}).
The parameters from Table~\ref{tab:par} are used in the form-factor parametrization,
and the full correlations are taken into account in the uncertainty of $d\Gamma/dq^2$.}
\label{tab:dGdq2}
\end{center}
\end{table}

\begin{figure}[!t]
  \begin{center}
    \begin{picture}(250,200)(0,0)
      \put(34,-10){\includegraphics[scale=0.3]{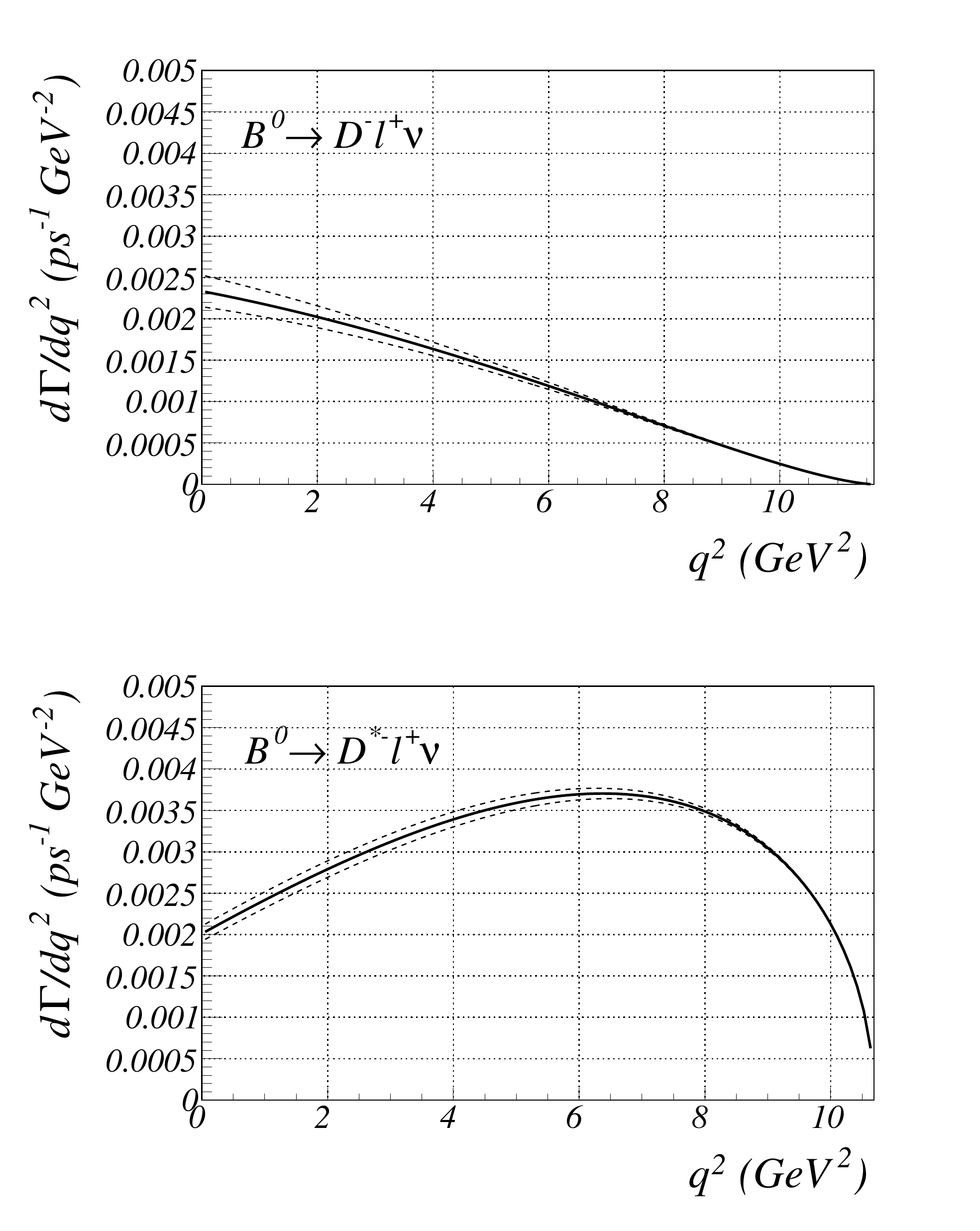}}
    \end{picture}
    \caption[dG/dq2]{\em $d\Gamma/dq^2$ for the form-factor parametrization of 
      Ref.~\cite{Caprini} and the HFAG 
      parameters as given in Table~\ref{tab:par}. 
      The uncertainty on $\rho^2$ is illustrated by the dotted curves.}
    \label{fig:dGdq2}
  \end{center}
\end{figure}

In order to determine the differential semi-leptonic decay rate at the appropriate momentum
transfer for the factorization test in (\ref{eq:a1}), we use the form-factor parametrization 
proposed by Caprini, Lellouch, and Neubert~\cite{Caprini}, with parameters summarized in
Table~\ref{tab:par}, yielding the rates shown in Fig.~\ref{fig:dGdq2}.

In the values of the semi-leptonic decay rates, the systematic uncertainty is estimated by propagating
the uncertainties from the parameters in Table~\ref{tab:par} to the appropriate 
value of $w$, taking the correlations into account.

\begin{table}[!b]
\small
\begin{center}
\begin{tabular}{|c|l|r|c c|c c|c c|} 
\hline
Topol. & Decay& BR\cite{PDG}&\multicolumn{4}{c|}{$|a_1(D_qP)|$}                         \\ \cline{4-7}
       &      & ($\times 10^{4}$)      &\multicolumn{2}{c|}{BaBar} &\multicolumn{2}{c|}{Belle}\\ 
\hline
$T'^{}$    &$\bar B^0_d\to D^{+}K^-$   &$2.0 \pm 0.6 $ &  0.89 &  $\pm$ 0.13 & 0.88 &  $\pm$ 0.16  \\
$T'^{*}$   &$\bar B^0_d\to D^{*+}K^-$  &$2.14\pm 0.16$ &  0.96 &  $\pm$ 0.05 & 0.99 &  $\pm$ 0.05  \\
\hline                                                                                             
$T+E$      &$\bar B^0_d\to D^{+}\pi^-$ & $26.8\pm 1.3$ &  0.88 &  $\pm$ 0.04 & 0.88 &  $\pm$ 0.09  \\
$T^*$+$E^*$&$\bar B^0_d\to D^{*+}\pi^-$& $27.6\pm 1.3$ &  0.98 &  $\pm$ 0.04 & 1.01 &  $\pm$ 0.04  \\
\hline
\end{tabular}
\caption[Factorization]
{\em Determination of the $|a_1(D_qP)|$ from the current data. The error is estimated 
by adding the uncertainties of the hadronic branching ratio and the semi-leptonic rate 
in quadrature. The correlations between the form-factor parameters for the 
semi-leptonic decay rate are taken into account. 
}\label{tab:a1}
\end{center}
\end{table}

\begin{figure}[!t]
  \begin{center}
    \begin{picture}(250,120)(0,0)
      \put(-30,-10){\includegraphics[bb=0 700 350 30,clip,scale=0.35,angle=90]{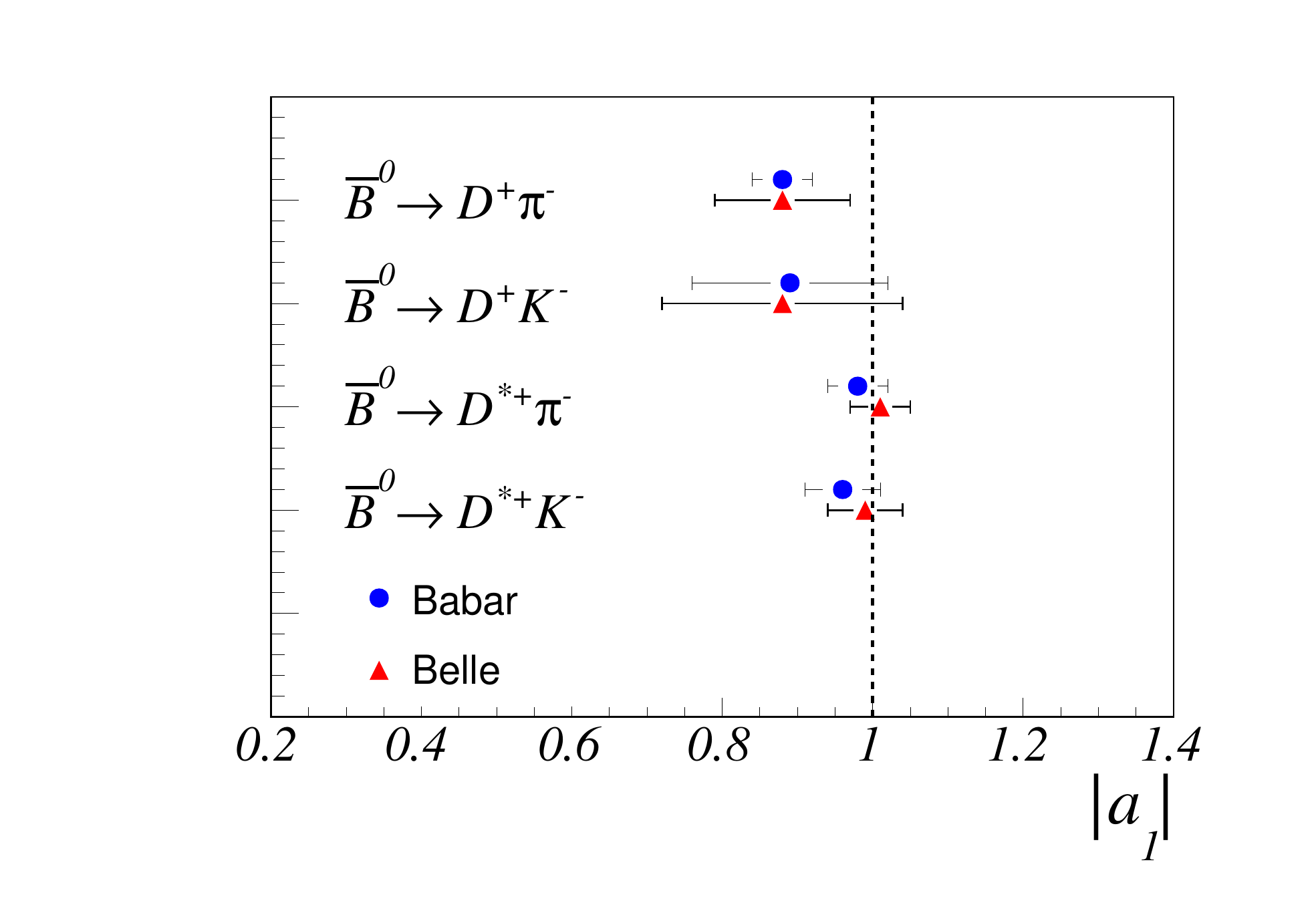}}
    \end{picture}
    \caption[a1]{\em The values for $|a_1(D_qP)|$ as obtained with the Belle and BaBar
      parameters from Table~\ref{tab:par}.
      The errors represent the error from the hadronic branching ratio~\cite{PDG}
      with the uncertainty of the semi-leptonic decay rate added in quadrature. 
      The full correlation matrix of the uncertainties in the determination of the form-factor
      parametrization of both the Belle and BaBar result is taken into account. 
      No uncertainty on the decay constants is included.}
    \label{fig:a1}
  \end{center}
\end{figure}

Using the numerical values from Table~\ref{tab:dGdq2} and the branching
ratios for the nonleptonic decays given by the PDG \cite{PDG}, we arrive at
the values for $|a_1(D_qP)|$ collected in Table~\ref{tab:a1} and compiled in
Fig.~\ref{fig:a1}. In naive factorization, we have $|a_1(D_qP)|=1$, while the 
QCD factorization analysis of Ref.~\cite{BBNS} results in an essentially 
universal value of $|a_1|\simeq 1.05$. It is interesting to note that the current 
experimental values of the $|a_1|$ favor a central value that is smaller than 
one, around $|a_1|\simeq 0.95$. Within the errors, we cannot resolve nonfactorizable
effects. Only the $\bar B^0_d \to D^+\pi^-$ decay shows a value of $a_1$ that is about $2\sigma$
away from factorization. 

We encourage the $B$ factories to determine the ratios of the semi-leptonic
differential decay rates and the relevant hadronic branching ratios directly. 
Correlated systematic uncertainties, such as the $D^{(*)}$ reconstruction efficiencies 
and the $D^{(*)}$ branching fractions, would cancel so that the
$B$-factory results could be fully exploited. 
These correlations are not considered in the errors estimated in Table~\ref{tab:a1}.

Recently, calculations became available that estimate electromagnetic
corrections to two-body $B$-meson decays into two light hadrons~\cite{Baracchini}.
They can be as large as 5\% for $B^0_d\to \pi^+\pi^-$ for a $E_{\gamma,max}=250$~MeV,
but we do not know to what extend these corrections are accounted for in 
the measurements of heavy-light decays.

As noted in Ref.~\cite{BBNS}, further tests of factorization are offered by the measurement
of the ratios of nonleptonic decay rates \cite{BBNS}:
\begin{eqnarray}
\lefteqn{\frac{\mbox{BR}(\bar B^0_d\to D^{+}\pi^-)}{\mbox{BR}(\bar B^0_d \to D^{*+} \pi^-)}}\nonumber\\
&&=\frac{(m_B^2 - m_D^2)^2 |\vec{q}|}{4m_B^2|\vec{q}|^3} 
\left( \frac{F_0(m_\pi^2)}{A_0(m_\pi^2)} \right)^2
\left| \frac{a_1(D\pi)}{a_1(D^*\pi)} \right|^2,
\end{eqnarray}
\begin{eqnarray}
\lefteqn{\frac{\mbox{BR}(\bar B^0_d\to D^{+}\rho^-)}{\mbox{BR}(\bar B^0_d \to D^{+} \pi^-)}}
\nonumber\\
&&=\frac{4m_B^2|\vec{q}|^3}{(m_B^2 - m_D^2)^2 |\vec{q}|}
\frac{f_\rho^2}{f_\pi^2}
\left( \frac{F_+(m_\rho^2)}{F_0(m_\pi^2)} \right)^2
\left| \frac{a_1(D\rho)}{a_1(D\pi)} \right|^2.
\end{eqnarray}
Using the branching ratios from Table~\ref{tab:a1} gives 
\begin{equation}
\left| \frac{a_1(D\pi)}{a_1(D^*\pi)} \right| 
 \frac{F_0(m_\pi^2)}{A_0(m_\pi^2)} = 0.95 \pm 0.03
\end{equation}
\begin{equation}
\left| \frac{a_1(D\rho)}{a_1(D\pi)} \right| 
 \frac{F_+(m_\rho^2)}{F_0(m_\pi^2)}  = 1.07 \pm 0.10,
\end{equation}
so that there is -- within the errors --  no evidence for any deviation from naive factorization.

It is worth noticing that in the case where the pseudoscalar is replaced by a
vector meson, the structure is much richer.  In this case factorization can be
tested through the longitudinal polarization of the $D^{*}$ mesons
\cite{FactTest-DPol}. This feature was exploited in the decay $\bar B^0_d\to
D^{*+}\rho^-$ in Ref.~\cite{FactCleo}, where the measured polarization was found
in excellent agreement with the factorization prediction within 2\%.  Other
final states such as $D^{*+}D^{*-}$, $D^{*+}D^{*-}_s$ and $D^{*+}\omega\pi^-$ 
further strengthen the agreement with factorization~\cite{FactBaBar}.

Finally, the large value for the longitudinal polarization of $B^0_s\to
D_s^{-*}\rho^+$ as reported in Ref.~\cite{FactBelle} not only agrees with
factorization, but is also -- within the errors -- in agreement with the value for 
$B_d\to D^{*-}\rho^+$, thereby supporting the $SU(3)$ flavor symmetry:
\begin{eqnarray}
f_L(B^0  \to D^{-*}  \rho^+) = & 0.885 & \pm 0.02     \\
f_L(B_s^0\to D_s^{-*}\rho^+) = & 1.05  & \pm 0.09.  
\end{eqnarray}
Here $f_L = \Gamma_L/\Gamma = |H_0|^2/(|H_{-1}|^2+|H_{0}|^2+|H_{+1}|^2)$.

\boldmath
\section{Information on $E$}\label{sec:E}
\unboldmath
Exchange topologies $E$ (see Fig.~\ref{fig:top}), which are naively expected to
be significantly suppressed with respect to the color-allowed $T$ amplitudes,
are examples where factorization is not expected to be a good approximation
\cite{BBNS}. In contrast to the $D^{(*)}K$ decays considered in the previous
section, the $\bar B^0_d\to D^{(*)+}\pi^-$ modes receive contributions from a
color-allowed tree and an exchange topology so that their decay amplitudes take
the following form:
\begin{equation}
\mbox{BR}(\bar B^0_d\to D^{(*)+}\pi^-) = |A(\bar B^0_d\to D^{(*)+}\pi^-)|^2 
\Phi^d_{D^{(*)}\pi}\tau_{B_d}, \nonumber
\end{equation}
where $\Phi^d_{D^{(*)}\pi}$ is a phase-space factor and
\begin{equation}
A(\bar B^0_d\to D^{(*)+}\pi^-)=T^{(*)}+E^{(*)}.
\end{equation}
The current experimental averages for their branching ratios are given 
in the lower half of Table~\ref{tab:a1}.

We will distinguish the $D^{(*)}\pi$ amplitudes from the $D^{(*)}K$ amplitudes
by the prime symbol. This will be relevant in Section~\ref{sec:su3}, where 
the validity of the $SU(3)$ flavor symmetry is further discussed.

The $E'^{(*)}$ amplitudes can actually be probed in three ways, namely by
comparing the hadronic branching fractions to the semi-leptonic decay rates as
was done in the previous section, by using the ratios of branching ratios
governed by the $T'^{(*)}$ and $T^{(*)}+E^{(*)}$ amplitudes, and by probing
$E'^{(*)}$ directly through the branching ratios of decays that originate only
from exchange topologies.

The comparison to the semi-leptonic rates is shown in the lower half of
Table~\ref{tab:a1}, and shows no sign of an enhancement of the $E'^{(*)}$
amplitudes with respect to the naive expectation \cite{BBNS}.

Let us next probe the $E'^{(*)}$ topologies through the ratios of branching
ratios, $\mbox{BR}(\bar B^0_d\to D^{(*)+}\pi^-)/\mbox{BR}(\bar B^0_d \to
D^{(*)+} K^-)$.  In the following, we will correct the $T'^{(*)}$ amplitudes
from $\bar B^0_d \to D^{(*)+} K^-$ for factorizable $SU(3)$-breaking
corrections, to allow for a direct comparison with the $T^{(*)}+E^{(*)}$
amplitude from $\bar B^0_d \to D^{(*)+}\pi^-$.  The factorizable
$SU(3)$-breaking corrections contain the pion and kaon decay constants $f_\pi$
and $f_K$, respectively, and the corresponding form factors, which we discussed
in the previous section:
\begin{equation}\label{T-SU3}
\left|\frac{T'^{(*)}}{T^{(*)}}\right|_{\mathrm{fact}} =
\left|\frac{V_{us}}{V_{ud}}\right|
\frac{f_K}{f_\pi}
\frac{F^{B\to D^{(*)}}(m^2_K)}{F^{B\to D^{(*)}}(m^2_\pi)}.
\end{equation}

In the case of the decays involving $D^{*+}$ mesons, the ratio of the branching
ratios has been measured with impressive precision \cite{RatioBaBar}: 
\begin{equation}
\frac{\mbox{BR}(\bar B^0_d\to D^{*+}K^-)}{\mbox{BR}(\bar B^0_d \to D^{*+} \pi^-)}
= (7.76\pm 0.34 \pm 0.29)\%,
\label{eq:BdDstarhRatio}
\end{equation}
which allows us to extract the ratio of $|T+E|$ and $|T'^{}|$ amplitudes. 
\begin{equation}\label{eq:TEoverTRatio}
\left|\frac{T'^{*}}{T^{*}+E^{*}}\right|
\overset{\mathrm{fact}}{\longrightarrow}
\left|\frac{T^{*}}{T^{*}+E^{*}}\right|
= 0.983 \pm 0.028.
\end{equation}
The consistency of the numerical value with 1 is remarkable and shows both a small
impact of the exchange topology and of nonfactorizable $SU(3)$-breaking effects.

Unfortunately, the $SU(3)$-counterpart  $\bar B^0_d \to D^{+} K^-$ of $\bar B^0_d\to D^{+}\pi^-$ 
still suffers from large uncertainties that are
introduced by the experimental value of $\mbox{BR}(\bar B^0_d \to D^{+} K^-)$,
yielding
\begin{equation}\label{rat-3}
\left|\frac{T'}{T+E}\right|
\overset{\mathrm{fact}}{\longrightarrow}
\left|\frac{T}{T+E}\right| 
= 0.99 \pm 0.15.
\end{equation}
The CDF collaboration has quoted the ratio
$\mbox{BR}(\bar B^0_d\to D^{+}K^-)/\mbox{BR}(\bar B^0_d \to D^{+} \pi^-)= 0.086 
\pm 0.005\,\mbox{(stat)}$ \cite{RatioCDF}, but has unfortunately not yet assigned a
systematic error. This result would lead to a numerical value of $1.07 \pm 0.03\,\mbox{(stat)}$
for the ratio in Eq.~(\ref{rat-3}). It would be interesting to get also an assessment of the 
corresponding systematic uncertainty.

Finally, we can also probe the exchange topologies directly through 
$\bar B^0_d\to D_s^{(*)+}K^-$ decays:
\begin{equation}
A(\bar B^0_d\to D_s^{(*)+}K^-)= E'^{(*)}.
\end{equation}
As in Eq.~(\ref{T-SU3}) we take differences in the final state into account through
\begin{equation}\label{E-SU3}
\left|\frac{E'^{(*)}}{E^{(*)}}\right|_{\mathrm{fact}} =
\frac{f_K f_{D_s^{(*)}}}{f_\pi f_{D_{}^{(*)}}},
\end{equation}
where the $f_{D^{(*)}}$ and $f_{D_s^{(*)}}$ are the decay constants of the
$D^{(*)} $ and $D_s^{(*)}$ mesons, respectively. In our numerical analysis, we use
$f_{D_s^{(*)}}/f_{D^{(*)}}=1.25\pm 0.06$  \cite{PDG}. The branching ratios are already well 
measured, as can be seen in Table~\ref{tab:e}~\cite{PDG}, and yield 
\begin{eqnarray}
\left|\frac{E}{T+E}\right|             & = & 0.073 \pm 0.006 \\ 
\left|\frac{E^{*}}{T^{*}+E^{*}}\right| & = & 0.066 \pm 0.006 ,
\label{eq:BdDhRatio2}
\end{eqnarray}
where we have rescaled the $E'^{}$ amplitude to the $E$ amplitude according
to Eq.~(\ref{E-SU3}).

\begin{figure}[!t]
  \begin{center}
    \begin{picture}(250,110)(0,0)
      \put(0,-10){\includegraphics[bb=10 250 525 525, scale=0.45]{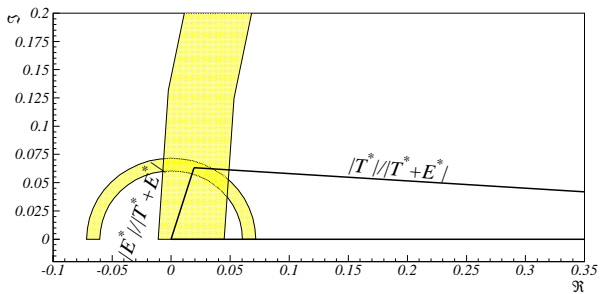}}
    \end{picture}
    \caption{\em The $E^{*}$, $T^{*}$ and $E^{*}+T^{*}$ amplitudes in the
      complex plane for the $B^0\to D_{(s)}^{*} \pi/K$ decays. The $E'^{*}$ and $T'^{*}$ amplitudes
      are rescaled by factorizable $SU(3)$ corrections 
      to match the $E^{*}$ and $T^{*}$ amplitudes for the  $B\to D^{*} \pi$
      case.}\label{fig:te}
  \end{center}
\end{figure}

It is instructive to illustrate the triangle relation between the $E^{(*)}$, $T^{(*)}$ and 
$E^{(*)}+T^{(*)}$ amplitudes in the complex plane. In Fig.~\ref{fig:te}, we show the 
situations emerging from the current data for the $B\to D^{*} P$ decays. 
While the $B\to D P$ decays still suffer from large uncertainties due to (\ref{rat-3}), we arrive 
at a significantly sharper picture for the $B\to D^{*} P$ modes. In particular, we can also 
determine the strong phase $\delta_*$ between the  $E'^{*}$ and $T'^{*}$ amplitudes,
which is given by $\delta_*\sim(77\pm 30)^\circ$. The favored large value of 
this phase explains the small impact of the $E^{*}$ amplitude on the total 
$\bar B^0_d\to D^{*+}\pi^-$ branching ratio. 


\begin{table}[!b]
\begin{center}
\begin{tabular}{|l|l|l|} 
\hline
 Decay              & meas. BR\cite{PDG} & pred. BR \\
                    & ($\times 10^{6}$)  &  ($\times 10^{6}$) \\
\hline
$\bar B^0_d\to D_s^{+}K^-$ & $30 \pm 4$         &                 \\
$\bar B^0_s\to D^{+}\pi^-$ &                    & $1.19\pm 0.16$  \\ 
\hline
$\bar B^0_d\to D_s^{*+}K^-$& $21.9 \pm 3$      &                 \\
$\bar B^0_s\to D^{*+}\pi^-$&                   & $0.90 \pm 0.12$ \\
\hline
\end{tabular}
\caption[Predict E$_s$]
{\em Predictions for the branching ratios of $B_s$ decays that occur only through
exchange topologies.}
\label{tab:e}
\end{center}
\end{table}

Other potentially interesting decays to obtain insights into the exchange topologies 
are the $\bar B^0_s\to D^{(*)+}\pi^-$ modes. Using the $U$-spin flavor symmetry, 
we expect
\begin{eqnarray}
\lefteqn{\frac{\mbox{BR}(\bar B^0_s\to D^{(*)+}\pi^-) }
              {\mbox{BR}(\bar B^0_d\to D^{(*)+}_s K^-)} }   \nonumber\\
& & =
\left|\frac{V_{us}}{V_{ud}}\right|^2
\left[\frac{f_{D^{(*)}}f_\pi f_{B_s}}{f_{D_s^{(*)}}f_K f_{B_d}}\right]^2 
\frac{\tau_{B_s}\Phi^s_{D^{(*)}\pi}}{\tau_{B_d}\Phi^d_{D_s^{(*)}K}}.\label{U-spin-test}
\end{eqnarray}
The predictions for the $B_s$ branching ratios using this relation are given in
Table~\ref{tab:e}. Unfortunately, it will be challenging for LHCb to measure this 
small branching ratio accurately since only a dozen of 
$\bar B^0_s\to D^{(*)+}\pi^-$ events are expected to be selected within the 1~fb$^{-1}$ 
data sample, which should be available by the end of 2011. However, for a luminosity 
of $(5\mbox{--}10)$~fb$^{-1}$, LHCb has the potential to discover these strongly suppressed 
decays. A future measurement of the ratios in Eq.~(\ref{U-spin-test})
would be an interesting probe of nonfactorizable $U$-spin-breaking effects.

\boldmath
\section{Isospin Triangles and Information on $C$}\label{sec:iso}
\unboldmath
The amplitudes for the three $B\to D^{(*)}\pi$ decays can be expressed in terms of
color-allowed and color-suppressed tree as well as exchange topologies. Alternatively,
the system can also be decomposed in terms of two isospin amplitudes, $A_{1/2}$ 
and $A_{3/2}$, which correspond to the transition into $D^{(*)}\pi$ final states with 
isospin $I=1/2$ and $I=3/2$, respectively~\cite{Rosner}. The ratio 
\begin{equation}\label{iso-rat}
\frac{A_{1/2}}{\sqrt{2} A_{3/2}}=1+{\cal O}(\Lambda_{\rm QCD}/m_b)
\end{equation}
is a measure of the departure from the heavy-quark limit~\cite{BBNS}, and has been measured by the 
CLEO~\cite{Cleo-iso} and BaBar collaborations \cite{BaBar-iso}. 

\begin{figure}[!t]
  \begin{center}
    \begin{picture}(250,210)(0,0)
      \put(-70,-30){\includegraphics[bb=10 180 525 545, clip=,scale=0.65]{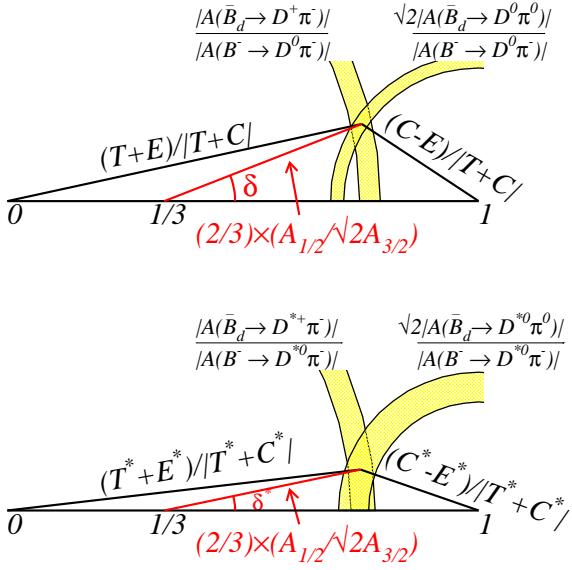}}
    \end{picture}
    \caption{\em Sketch of the $T+E$ and $C-E$ amplitudes, normalized to $|T+C|$ (and corrected
      for differences in phase space) in the
      complex plane for the $B\to D \pi$ decays (top) and $B\to D^{*} \pi$ decays (bottom).
      The ratio of isospin amplitudes in Eq.~(\ref{iso-rat-calc}) is also drawn.}\label{fig:iso}
  \end{center}
\end{figure}

Using updated information on the nonleptonic branching ratios, we will repeat this 
isospin analysis. The corresponding isospin relations read as
\begin{equation}
A(\bar B^0_d\to D^+\pi^-)=\sqrt{\frac{1}{3}}A_{3/2}+\sqrt{\frac{2}{3}}A_{1/2}=T+E 
\end{equation}
\begin{equation}
\sqrt{2}A(\bar B^0_d\to D^0\pi^0)=\sqrt{\frac{4}{3}}A_{3/2}-\sqrt{\frac{2}{3}}A_{1/2}=C-E
\end{equation}
\begin{equation}
A(B^- \to D^0\pi^-)=\sqrt{3}A_{3/2}=T+C,
\end{equation}
so that 
\begin{equation}
A_{1/2}=\frac{2T-C+3E}{\sqrt{6}}
\label{eq:a12}
\end{equation}
\begin{equation}
A_{3/2}=\frac{T+C}{\sqrt{3}},
\label{eq:a32}
\end{equation}
which leads to the following expression,
\begin{equation}\label{iso-rat-calc}
\frac{A_{1/2}}{\sqrt{2}A_{3/2}}=1-\frac{3}{2}\left(\frac{C-E}{T+C} \right).
\end{equation}
The $(T+E)$, $(C-E)$ and $(T+C)$ amplitudes can be depicted in the complex plane,
and related to the ratio of isospin amplitudes, as shown in Fig.~\ref{fig:iso}.

The absolute values of the amplitudes $A_{1/2}$ and $A_{3/2}$
can also be obtained directly from the measured decay rates:
\begin{eqnarray}
|A_{1/2}|^2 & = & |A(D^+\pi^-)|^2 + |A(D^0\pi^0)|^2 -  \frac{1}{3} |A(D^0\pi^-)|^2 \nonumber \\
|A_{3/2}|^2 & = & \frac{1}{3} |A(D^0\pi^-)|^2,
\end{eqnarray}
which can be expressed in terms of partial decay widths through
$|A(D\pi)|^2 = \Gamma(D\pi)/\Phi^d_{D\pi}$, i.e. 
corrected for the small differences in phase space, which mainly leads to a small but measurable 
correction for $D^0\pi^0$ final state.
The relative strong phase between the $I=3/2$ and $I=1/2$ amplitudes can be calculated with
\begin{equation}
\cos \delta = \frac{3|A(D^+\pi^-)|^2 + |A(D^0\pi^-)|^2 - 6 |A(D^0\pi^0)|^2}{6
\sqrt{2}|A_{1/2}||A_{3/2}|},
\end{equation}
and similar for $\delta^{*}$.

We find the following numerical results:
\begin{eqnarray}
\left|\frac{A_{1/2}}{\sqrt{2}A_{3/2}}\right|_{D\pi}    & = & 0.676 \pm 0.038 \\
\left|\frac{A_{1/2}}{\sqrt{2}A_{3/2}}\right|_{D^*\pi}  & = & 0.639 \pm 0.039,
\end{eqnarray}
which are complemented by 
\begin{eqnarray}
\cos \delta     & = & 0.930  ^{+0.024}_{-0.022}  , \\
\cos \delta^{*} & = & 0.979  ^{+0.048}_{-0.043}  .
\end{eqnarray}
The nominal value is calculated from the central values of the 
branching fractions, whereas the $\pm 1\sigma$ confidence interval
is defined as the integral of 68.3\% of the total area 
of its likelihood function, similar to the procedure followed
in Ref.~\cite{BaBar-iso}. The corresponding central values for the strong
phases then become $\delta = 21.6^\circ$ and 
$\delta^{*} = 11.9^\circ$ for the $D\pi$ and $D^*\pi$ case,
respectively. 

Comparing with (\ref{iso-rat}), we observe that the isospin-amplitude ratio shows 
significant deviations from the heavy-quark limit. In view of our analysis of the exchange
topologies in Section~\ref{sec:E} and the expression in (\ref{iso-rat-calc}) we can trace
this feature back to the color-suppressed $C$ topologies.

\boldmath
\section{Tests of $SU(3)$ Symmetry}\label{sec:su3}
\unboldmath
Let us next probe the impact of nonfactorizable $SU(3)$-breaking corrections
in $B$-meson decays into heavy-light final states. To this end, we compare the 
three $B\to D^{(*)}\pi$ decays with their $SU(3)$-related $B\to D_{(s)}^{(*)}K$ channels,
which have decay amplitudes of the following structure:
\begin{equation}
A(\bar B^0_d \to D^{0}K^0)   = C'
\end{equation}
\begin{equation}
A(\bar B^0_d \to D_s^{+}K^-) = E'
\end{equation}
\begin{equation}
A(\bar B^0_d \to D^{+}K^-)   = T'
\end{equation}
\begin{equation}
A(B^- \to D^{0}K^-)          = T' + C'.
\end{equation}
Here the notation is as above and the primes remind us again that we are
dealing with $b\to c\bar c s$ quark-level transitions in this case. In order to  quantify 
the validity of the $SU(3)$ flavor symmetry, we can perform the following four
experimental tests:
\begin{itemize}
\item[(i)]   Consistency between $E'^{*}$,  $T'^{*}$ and $T^{*}+E^{*}$;
\item[(ii)]  Consistency between $E'^{(*)}$, $C'^{(*)}$ and $C^{(*)}-E^{(*)}$;
\item[(iii)] Ratio of $|T^{(*)}+C^{(*)}|$ and $|T'^{(*)}+C'^{(*)}|$; 
\item[(iv)]  Prediction for $E'^{(*)}$, based on all amplitudes listed in Table~\ref{tab:tec}
apart from $A(\bar B^0_d \to D_s^{+}K^-)$.
\end{itemize} 
Tests (ii--iv) can be performed with both the $B\to D_{(s)}P$ and the $B\to D_{(s)}^*P$ systems.
On the other hand, due to the large uncertainty affecting $\mbox{BR}(\bar B^0_d \to D^{+} K^-)$,
test (i) can currently only be applied to the $D^*$ case. 
We will use the values for the branching fractions as listed in Table~\ref{tab:tec}.
The size of the $E'^{*}$, $T'^{*}$ and
$T^{*}+E^{*}$ amplitudes are internally consistent, as is shown by the overlapping circles 
in Fig.~\ref{fig:iso}. As we noted already in Section~\ref{sec:E}, this also indicates that there 
are no large nonfactorizable $SU(3)$-breaking effects in the $E'^{*}$ or $T'^{*}$ amplitudes.

\begin{figure}[!t]
  \begin{center}
    \begin{picture}(250,210)(0,0)
      \put(-70,-30){\includegraphics[bb=10 180 525 545, clip=,scale=0.65]{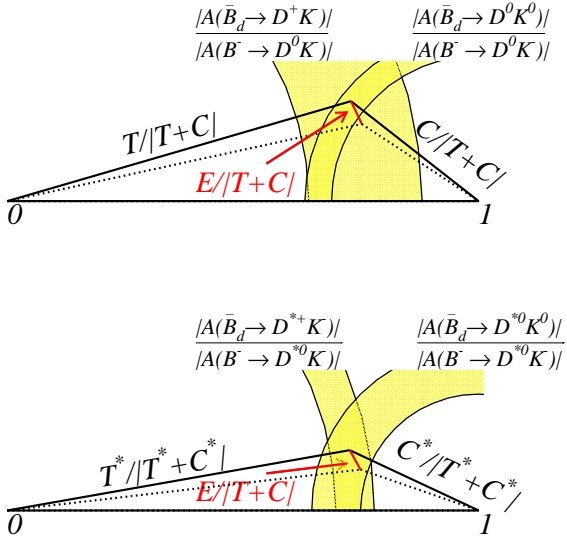}}
    \end{picture}
    \caption{\em Sketch of the $T'$ and $C'$ amplitudes, normalized to $|T'+C'|$ (and corrected
      for factorizable SU(3)-breaking effects) in the
      complex plane for the $B\to D K$ decays (top) and $B\to D^{*}K$ decays (bottom).
      The predicted $E$ amplitude, assuming SU(3) symmetry is also drawn.}\label{fig:E}
  \end{center}
\end{figure}

Similarly to Eq.(\ref{eq:TEoverTRatio}) we can check the consistency between the 
$E'^{(*)}$, $C'^{(*)}$ and $C^{(*)}-E^{(*)}$ amplitudes.
As before we will correct the $C'^{(*)}$ amplitudes from $\bar B^0_d \to D^{(*)0} K^0$ 
for factorizable $SU(3)$-breaking corrections, to allow for a direct comparison with the
$C-E$ amplitude,
\begin{equation}\label{C-SU3}
\left|\frac{C'^{(*)}}{C^{(*)}}\right|_{\mathrm{fact}} =
\left|\frac{V_{us}}{V_{ud}}\right|
\frac{F^{B\to K}(m^2_{D^{(*)}})}{F^{B\to \pi}(m^2_{D^{(*)}})},
\end{equation}
where we use the parametrization for $F^{B\to \pi/K}$ from Ref.~\cite{Ftopi-Ball}.
We extract the following ratio of $|C-E|$ and $|C'^{}|$ amplitudes:
\begin{eqnarray}\label{eq:CEoverERatio}
\left|\frac{C-E}{C}            \right|   & = & 0.913 \pm 0.074 \\
\left|\frac{C^{*}-E^{*}}{C^{*}}\right|   & = & 0.89 \pm 0.18,
\end{eqnarray}
where the factorizable $SU(3)$-breaking corrections are taken into account.
Again, the ratio is close to 1, indicating that the contribution of $E$ is small,
and that there are no unexpected nonfactorizable $SU(3)$ violating effects, in addition to the
factorizable $SU(3)$ corrections. 
This is remarkable in view of the nonfactorizable character of the color-suppressed 
decays.

A direct measure of the size of $SU(3)$-breaking effects in $B\to D^{(*)}P$ decays
is provided by the ratio of the $|T^{(*)}+C^{(*)}|$ and $|T'^{(*)}+C'^{(*)}|$ amplitudes:
\begin{equation}
\frac{\mbox{BR}(\bar B^-\to D^{(*)0}\pi^-)}{\mbox{BR}(\bar B^-\to D^{(*)0} K^-)}  =
\left| \frac{T^{(*)}+C^{(*)}}{T'^{(*)}+C'^{(*)}} \right|^2
\frac{\Phi_{D^{(*)}\pi}}{\Phi_{D^{(*)}K}},
\label{SU3-test2}
\end{equation}
where the ratio of branching ratios has been measured for the $D^0$ case as
follows \cite{BtoD0KBelle}:
\begin{equation}
\frac{\mbox{BR}(B^-\to D^0K^-)}{\mbox{BR}(B^- \to D^0 \pi^-)}
= (7.7\pm 0.5 \pm 0.6)\%.
\label{eq:BtoD0KRatio}
\end{equation}
If we include factorizable $SU(3)$-breaking effects through the corresponding decay
constants and form factors, the numerical values of the relevant amplitude ratios
are given as follows:
\begin{equation}
\left| \frac{T    +C}    {T'^{} +C'^{} } \right|
\left|\frac{V_{us}}{V_{ud}}\right|
\frac{f_K}{f_\pi}
\frac{F^{B\to D}(m^2_K)}{F^{B\to D}(m^2_\pi)}
  =  0.997 \pm 0.047 ,                        
\end{equation}
\begin{equation}
\left| \frac{T^{*}+C^{*}}{T'^{*}+C'^{*}} \right| 
\left|\frac{V_{us}}{V_{ud}}\right|
\frac{f_K}{f_\pi}
\frac{F^{B\to D^{*}}(m^2_K)}{F^{B\to D^{*}}(m^2_\pi)}
 =  0.995 \pm 0.048 .                            
\end{equation}
The factorizable $SU(3)$-breaking effects for the $C$ amplitudes (\ref{C-SU3}) 
are numerically close to the ones for the $T$ amplitudes (\ref{T-SU3}),
and since the $T$ amplitude is the dominant amplitude here, we rescale in the same way as
in Eq.~(\ref{T-SU3}).

The consistency with 1 is remarkable. In particular, we find that nonfactorizable 
$SU(3)$-breaking effects are smaller than 5\%, even in decays that have a large 
contribution from color-suppressed amplitudes where factorization does not work
well, as we have seen in the previous section.

Finally, we can -- in analogy to Fig.~\ref{fig:iso} --  construct a second amplitude triangle, 
which involves now the $T'$ and $C'$ amplitudes, as shown in Fig.~\ref{fig:E}.
If we rescale the primed amplitudes involving a kaon in the final state to the 
amplitudes with a pion in the final state by correcting for the 
factorizable $SU(3)$-breaking corrections, the distance between the apexes
of Figs.~\ref{fig:iso} and~\ref{fig:E} shows graphically how $|E|$ can be predicted.
The consistency between the corresponding value and the measured value for $|E'|$ 
from  $\mbox{BR}(\bar B^0_d\to D_s^-\pi^+)$ is a direct probe for nonfactorizable 
$SU(3)$-breaking effects in nonleptonic decays of the type $B\to D^{(*)}P$.
The numerical picture for the $D\pi$ and $D^{*}\pi$ cases is 
still not precise enough to predict the measured value:
\begin{equation}
\left|\frac{E}{T+C}\right|_{\mathrm{meas}} =  0.056 \pm 0.004
\end{equation}
and 
\begin{equation}
\left|\frac{E^{*}}{T^{*}+C^{*}}\right|_{\mathrm{meas}}  =  0.047 \pm 0.004,
\end{equation}
respectively, where $E'^{(*)}$ is rescaled to $E^{(*)}$ according to Eq.~(\ref{E-SU3}).
The knowledge of $T'^{(*)}$ and $C'^{(*)}$ will probably be improved
in the near future, which will provide another interesting test of 
the validity of the $SU(3)$ flavor symmetry.

In the factorization tests discussed above, we did not consider $B_s$ decays. In this context,
interesting information on $SU(3)$-breaking effects can be obtained from the comparison 
between the polarization amplitudes of $B_s\to J/\psi \phi$ and $B_d\to J/\psi K^{*0}$ decays,
which are found in excellent agreement with one another \cite{BspsiphiCDF,BspsiphiD0}. Moreover, also
analyses of $B_s\to K^+K^-$ and $B_s\to \pi^\mp K^\pm$ modes and their comparison with
$B\to \pi K, \pi\pi$ decays do not show any indications of large nonfactorizable $SU(3)$-breaking
corrections \cite{BsKK}. A similar comment applies to the $B^0_s\to J/\psi K_{\rm S}$ mode
\cite{BFK}, which has recently been observed by the CDF collaboration \cite{BspsiKS}.

\begin{table}[!t]
\begin{center}
\begin{tabular}{|c|l||r|r|} 
\hline
Topology                             & Final state    & \multicolumn{2}{c|}{meas. BR ($\times 10^{4}$) \cite{PDG}}  \\
                                     &                &$\bar B^0_d\to DP$ & $\bar B^0_d\to D^*P$  \\
\hline						       						       
\hline						       						       
\multicolumn{4}{|c|}{Isospin amplitudes}                                                          \\
\hline						       					      	 
$T^{(*)} + E^{(*)}$                  & $D^{(*)+}\pi^-$&$26.8 \pm 1.3$ & $27.6 \pm 1.3 $           \\
$\frac{1}{\sqrt{2}}(C^{(*)}-E^{(*)})$& $D^{(*)0}\pi^0$&$2.61 \pm 0.24$& $1.7  \pm 0.4 $           \\
$T^{(*)} + C^{(*)}$                  & $D^{(*)0}\pi^-$&$48.4 \pm 1.5$ & $51.9 \pm 2.6 $           \\
\hline						       					      	 
\multicolumn{4}{|c|}{Amplitudes used to probe $SU(3)$ symmetry}                                   \\
\hline						       					      	 
$E'^{(*)}$                           & $D_s^{(*)+}K^-$&$0.30 \pm 0.04$& $0.219\pm 0.03$           \\
$C'^{(*)}$                           & $D^{(*)0}K^0$  &$0.52 \pm 0.07$& $0.36 \pm 0.12$           \\
$T'^{(*)}$                           & $D^{(*)+}K^-$  &$2.0  \pm 0.6$ & $2.14 \pm 0.16$           \\
$T'^{(*)}+C'^{(*)}$                  & $D^{(*)0}K^-$  &$3.68 \pm 0.33$& $4.21 \pm 0.35$           \\
\hline
\end{tabular}
\caption[Predict C$_s$]
{\em Branching fractions used in the various tests of the SU(3) flavor symmetry.}
\label{tab:tec}
\end{center}
\end{table}

%
%
%
\boldmath
\section{Application: Extraction of $f_d/f_s$}\label{sec:appl}\unboldmath
As we have seen in Section~\ref{sec:E}, the impact of the exchange topology on the 
$\bar B^0_d\to D^{*+}\pi^-$ decays is small. Consequently, this channel looks at first
sight also interesting for another implementation of the method for the determination of 
$f_d/f_s$ at LHCb proposed by us in Ref.~\cite{FST}. Here we have to compare it with the 
$\bar B^0_s\to D_s^{*+}\pi^-$ channel. Unfortunately, the reconstruction of the $D_s^{*+}$ is 
challenging at LHCb. However, the $\bar B^0_d\to D^{+}\pi^-$ and $\bar B^0_s\to D_s^{+}\pi^-$
modes are nicely accessible at this experiment. In view of the similar patterns of the modes 
involving $D^{*}$ and $D$ mesons discussed above, we expect that also in the 
$\bar B^0_d\to D^{+}\pi^-$ channel the exchange amplitude plays a minor role. The
expression for the extraction of $f_d/f_s$ from these channels reads as follows:
\begin{equation}\label{fs-det}
\frac{f_d}{f_s}=1.018\, \frac{\tau_{B_s}}{\tau_{B_d}}
\left[\tilde{\cal N}_a {\cal N}_F {\cal N}_E
\frac{\epsilon_{D_s \pi}}{\epsilon_{D_d \pi}}
\frac{N_{D_d \pi}} {N_{D_s\pi}}\right],
\end{equation} 
where the numerical factor takes phase-space effects into account,
\begin{equation}\label{NF_def-2}
 \tilde{\cal N}_a \equiv \left|\frac{a_1(D_s\pi)} {a_1(D_d\pi)}\right|^2,
 \quad {\cal N}_F \equiv \left[\frac{F_0^{(s)}(m_\pi^2)}{F_0^{(d)}(m_\pi^2)}\right]^2,
\end{equation}
describe $SU(3)$-breaking effects, and
\begin{equation}\label{NE_def-2}
 \quad {\cal N}_E \equiv \left|\frac{T}{T+E}\right|^2
\end{equation}
takes into account the effect of the exchange diagram, which was absent
in the $\bar B^0_d\to D^{+}K^-$ decay~\cite{FST}.

The difference of $|a_1|$ from unity at the order of 5\% discussed in Section~\ref{sec:T}
leads to an uncertainty of about $10\%$ on the theoretical prediction of the hadronic 
branching ratio. Assuming an $SU(3)$ suppression in the ${\cal N}_a$ factor  introduced 
in Ref.~\cite{FST} and the $\tilde{\cal N}_a$ by a factor $\sim 5$, which is still generous 
in view of the analysis of the $SU(3)$-breaking effects in Section~\ref{sec:su3}, we 
arrive at an uncertainty of about 2\% for ${\cal N}_a$ and $\tilde{\cal N}_a$.
This experimentally constrained error is fully consistent with the theoretical discussion
given in Ref.~\cite{FST}. 

Unfortunately, the $B_s\to D_s$ form factors entering ${\cal N}_F$ have so far received 
only small theoretical attention. In Ref.~\cite{chir}, such effects were explored using 
heavy-meson chiral perturbation theory, while QCD sum-rule techniques were applied 
in Ref.~\cite{BCN}. The numerical value given in the latter paper yields 
${\cal N}_F = 1.24 \pm 0.08$.

Finally, in contrast to the determination of $f_s/f_d$ by means of the 
$\bar B^0_d\to D^{+}K^-$, $\bar B^0_s\to D_s^+\pi^-$ system \cite{FST}, 
we have to deal with the ${\cal N}_E$ factor in (\ref{fs-det}). Using (\ref{eq:TEoverTRatio}) and 
adding an additional 5\% uncertainty to account for possible differences
between the $D$ and $D^{*}$ cases, we obtain 
${\cal N}_E = 0.966 \pm 0.056 \pm 0.05$.

Interestingly, the CDF collaboration has already published the
ratio~\cite{BtoDpiCDF}:
\begin{equation}
\frac{\epsilon_{D_d \pi}}{\epsilon_{D_s \pi}}
\frac{N(D^-_s(\phi \pi^-)\pi^+)}{N(D^-(K^+\pi^-\pi^-)\pi^+)} = 0.067 \pm 0.04.
\end{equation}
After taking the branching fractions of the $D$-mesons 
into account, $\mbox{BR}(D^-\to K^+\pi^-\pi^-)=(9.4 \pm 0.4)\%$ and
 $\mbox{BR}(D_s^-\to \phi\pi^-)=(2.32 \pm 0.14)\%$, we obtain: 
\begin{equation}\label{eq:CDFB2Dpi}
\frac{\epsilon_{D_d \pi}}{\epsilon_{D_s \pi}}
\frac{N_{D_s\pi}}{N_{D_d\pi}} = 0.271 \pm 0.016 \pm 0.020 ~(\mbox{BR}(D)).
\end{equation}
If we use now Eq.~(\ref{fs-det}), we can convert this number into a value of $f_s/f_d$. 
Assuming $\tilde{\cal N}_a = 1.00 \pm 0.02$ and ${\cal N}_E = 0.966 \pm 0.056$, we obtain
the nonleptonic result
\begin{equation}\label{fsfd-hadronic}
\left( \frac{f_s}{f_d}\right)_{\mathrm{NL}} = 0.285 \pm 0.036,
\end{equation}
for ${\cal N}_F=1$, where all errors have been added in quadrature. 
Here we have a theoretical error of 8.2\% on top of an experimental error of 9.4\% from 
(\ref{eq:CDFB2Dpi}) and $\tau_{B_s}/\tau_{B_d} = 0.965 \pm 0.017$. As discussed
in Ref.~\cite{FST}, we expect ${\cal N}_F \geq 1$, which may result in a decrease of 
${f_s}/{f_d}$. Lattice results of the form-factor ratio entering ${\cal N}_F$ will hopefully 
be available soon. In order to surpass the possible future experimental uncertainty,
knowledge on the corresponding $SU(3)$-breaking corrections would be needed at the $20\%$ level.

It is interesting to compare the result in (\ref{fsfd-hadronic}) with the published ratio 
of fragmentation functions extracted from semi-inclusive $\bar B \to D \ell^- \bar \nu_\ell X$ 
decays \cite{bib:CDF_fdfs3}. The reconstructed $D \ell^-$ signal yields are related to the 
number of produced $b$ hadrons by assuming the $SU(3)$ flavor symmetry and neglecting
$SU(3)$-breaking corrections  (i.e.\ 
\ $\Gamma(\bar{B}_d^0\to \ell^-\bar{\nu}_\ell D^+)=\Gamma(\bar{B}^0_s\to \ell^-\bar{\nu}_\ell D_s^+)$, 
which corresponds to ${\cal N}_F = 1$). Together with an earlier result using double 
semi-leptonic decays (containing two muons and either a $K^*$ or a 
$\phi$ meson)~\cite{bib:CDF_fdfs1}, the average value $f_s/(f_d+f_u) = 0.142 \pm 0.019$ 
was obtained~\cite{PDG}, which can be written as
\begin{equation}\label{fsfd-leptonic}
\left(\frac{f_s}{f_d}\right)_{\rm{SL}} = 0.284 \pm 0.038.
\end{equation}
The consistency of this result with (\ref{fsfd-hadronic}) is remarkable.
Note that the uncertainties on the $\mbox{BR}(D_{(s)})$
and $D_{(s)}$-meson reconstruction efficiencies are expected to be correlated 
in (\ref{fsfd-hadronic}) and (\ref{fsfd-leptonic}).

As the mode $\bar B^0_d\to D^{+}\pi^{-}$ is Cabibbo-favored with 
respect to the $\bar B^0_d\to D^{+}K^{-}$ channel, it allows an early measurement of the 
hadronization fraction at LHCb. Moreover, the possibility of using $\bar B^0_d\rightarrow D^{+}\pi^{-}$ 
offers a useful experimental handle to keep systematic uncertainties due to particle identification
under control. These decays can be reconstructed with $D^{+}\to K^-\pi^+\pi^+$ and 
$D_s^+ \to K^+ K^- \pi^+$ final states. The uncertainty on the ratio of the two efficiencies 
$\epsilon_{D_s\pi}/\epsilon_{D\pi}$ is expected to be small since the topology is the same 
and the main difference is the particle identification of one of the kaons coming from the charmed 
meson. The number of events in $10\,\mbox{pb}^{-1}$~\cite{bib:CKM2010} is expected to 
be $\sim 3000$ for $\bar B^0_d\to D^+\pi^-$ and about $10\times$ smaller (depending on the value of 
$f_d/f_s$) for $\bar B^0_s\to D_s^{+}\pi^-$,
with $D^+\to K^-\pi^+\pi^+$ and $D_s^+\to K^-K^+\pi^+$, respectively.
Therefore this would allow LHCb to make a precise 
measurement of $f_d/f_s$ with a few tens $\mbox{pb}^{-1}$.

\section{Conclusions}\label{sec:concl}
We have considered nonleptonic $B$-meson decays of the kind $B\to D^{(*)}_{(s)} P$ 
and have performed tests of how well these channels are described by factorization
and $SU(3)$ flavor-symmetry relations. Using data from semi-leptonic $B$ decays
to determine the relevant $B \to D^{(*)}$ form factors, we could not resolve nonfactorizable
effects within the current experimental precision, which is as small as about $5\%$ in
the most fortunate cases. Using data on nonleptonic decays to probe exchange topologies, 
we obtained a picture with amplitudes as naively expected, i.e.\ without any enhancement due 
to long-distance effects. However, in an isospin analysis of the $B\to D^{(*)}\pi$ system, 
we found significant corrections to the heavy-quark limit, which could be traced back to
nonfactorizable contributions to color-suppressed tree contributions. Concerning the 
$SU(3)$ flavor symmetry, we did not find any indication for nonfactorizable
$SU(3)$-breaking corrections, with a resolution as small as $5\%$. 

These results support -- from an experimental point of view -- the intrinsic theoretical errors 
for a determination of the ratio $f_s/f_d$ of fragmentation functions from a simultaneous 
measurement of the $\bar B^0_d\to D^+K^-$ and $\bar B^0_s\to D_s^+\pi^-$ modes, as
proposed and discussed in Ref.~\cite{FST}. 

We found an interesting variant of 
this method, which arises if we replace the $\bar B^0_d\to D^+K^-$ channel by 
$\bar B^0_d\to D^+\pi^-$. In this case, we have then also to deal with a contribution 
from an exchange topology, which we constrain experimentally. Interestingly, the 
CDF collaboration has already published a ratio of the corresponding event numbers, 
which we can convert into $(f_s/f_d)_{\rm NL}=0.285\pm0.036$, 
with a smaller error than and in excellent agreement with 
the result from analyses of semi-leptonic decays at CDF.
It should be noted that in these values $SU(3)$-breaking effects in the ratio of the relevant
$B\to D$ and $B_s\to D_s$ form factors were neglected, which could reduce the value
of $f_s/f_d$. In the near future, precise lattice calculations of this quantity should become 
available. The extraction of $f_s/f_d$ from the $\bar B^0_s\to D_s^+\pi^-$,
$\bar B^0_d\to D^+\pi^-$ system, as proposed in this paper, is interesting for the early 
data taking at LHCb.

\noindent{\it Acknowledgments} \\
We would like to thank Barbara Storaci for many valuable discussions. 
This work is supported by the Netherlands Organisation for Scientific Research (NWO).

\newpage

\begin{thebibliography}{99}
%
%
%
\bibitem{fact}D.~Fakirov and B.~Stech,
  Nucl.\ Phys.\  B {\bf 133}, 315 (1978);
   N.~Cabibbo and L.~Maiani,
  Phys.\ Lett.\  B {\bf 73}, 418 (1978)
  [Erratum-ibid.\  B {\bf 76}, 663 (1978)].
  
\bibitem{BGR}A.~J.~Buras, J.-M.~G\'erard and R.~R\"uckl,
  Nucl.\ Phys.\  B {\bf 268}, 16 (1986).
  
\bibitem{bjor}J.~D.~Bjorken,
  Nucl.\ Phys.\ Proc.\ Suppl.\  {\bf 11}, 325 (1989).
  
\bibitem{DG}M.~J.~Dugan and B.~Grinstein,
  Phys.\ Lett.\  B {\bf 255}, 583 (1991).
  
\bibitem{BBNS} M.~Beneke, G.~Buchalla, M.~Neubert and C.~T.~Sachrajda,
  Nucl.\ Phys.\  B {\bf 591}, 313 (2000)
  [arXiv:hep-ph/0006124].

\bibitem{SCET}C.~W.~Bauer, D.~Pirjol and I.~W.~Stewart,
  Phys.\ Rev.\ Lett.\  {\bf 87}, 201806 (2001)
  [arXiv:hep-ph/0107002].

\bibitem{GHLR}M.~Gronau, O.~F.~Hernandez, D.~London and J.~L.~Rosner,
  Phys.\ Rev.\  D {\bf 52}, 6356 (1995)
  [arXiv:hep-ph/9504326].
  
\bibitem{BS}A.~J.~Buras and L.~Silvestrini,
  Nucl.\ Phys.\  B {\bf 569}, 3 (2000)
  [arXiv:hep-ph/9812392].

\bibitem{FST}R.~Fleischer, N.~Serra and N.~Tuning,
  Phys.\ Rev.\  D {\bf 82}, 034038 (2010)
  [arXiv:1004.3982 [hep-ph]].
  
\bibitem{PDG}K.~Nakamura {\it et al.}  [Particle Data Group],
  J.\ Phys.\ G {\bf 37}, 075021 (2010).

\bibitem{BStone}D.~Bortoletto and S.~Stone,
  Phys.\ Rev.\ Lett.\  {\bf 65}, 2951 (1990).
 
\bibitem{Caprini} I.~Caprini, L.~Lellouch and M.~Neubert,
  Nucl.\ Phys.\  B {\bf 530}, 153 (1998)
  [arXiv:hep-ph/9712417].

\bibitem{SemilepBelleDst}
  W.~Dungel {\it et al.}  [Belle Collaboration],
  Phys.\ Rev.\  D {\bf 82}, 112007 (2010)
  [arXiv:1010.5620 [hep-ex]].

\bibitem{bib:HFAG}D. Asner {\it et al.}\ [Heavy Flavor Averaging Group],
  arXiv:1010.1589 [hep-ex].

\bibitem{SemilepBelleD}
  K.~Abe {\it et al.}  [Belle Collaboration],
  Phys.\ Lett.\  B {\bf 526} (2002) 258
  [arXiv:hep-ex/0111082].

\bibitem{SemilepBaBarD}
  B.~Aubert {\it et al.}  [BABAR Collaboration],
  Phys.\ Rev.\ Lett.\  {\bf 104} (2010) 011802
  [arXiv:0904.4063 [hep-ex]].

\bibitem{SemilepBaBarDst}
  B.~Aubert {\it et al.}  [BaBar Collaboration],
  Phys.\ Rev.\  D {\bf 77}, 032002 (2008)
  [arXiv:0705.4008 [hep-ex]].

  \bibitem{Baracchini}
  E.~Baracchini and G.~Isidori,
  Phys.\ Lett.\  B {\bf 633} (2006) 309
  [arXiv:hep-ph/0508071].

\bibitem{FactTest-DPol}C. Reader and N. Isgur, Phys.\ Rev.\ D {\bf 47}, 1007 (1993);
Z.~Ligeti {\it et al.}, Phys.\ Lett.\ {\bf B507}, 142 (2001).

\bibitem{FactCleo}
  S.~E.~Csorna {\it et al.}  [CLEO Collaboration],
  Phys.\ Rev.\  D {\bf 67} (2003) 112002
  [arXiv:hep-ex/0301028].
  
\bibitem{FactBaBar}B. Aubert {\it et al.}\ [BaBar Collaboration],  
Phys.\ Rev.\  D {\bf 74}, 012001 (2006)
[arXiv:hep-ex/0604009].

\bibitem{FactBelle}
  R.~Louvot {\it et al.}  [Belle Collaboration],
  Phys.\ Rev.\ Lett.\  {\bf 104}, 231801 (2010)
  [arXiv:1003.5312 [hep-ex]].

\bibitem{RatioBaBar}B. Aubert {\it et al.}\ [BaBar Collaboration],  
Phys.\ Rev.\ Lett.\  {\bf 96}, 011803 (2006)
[arXiv:hep-ex/0509036].

\bibitem{RatioCDF}
  T.~Aaltonen {\it et al.}  [CDF Collaboration],
  Phys.\ Rev.\ Lett.\  {\bf 103} (2009) 191802
  [arXiv:0809.0080 [hep-ex]].

\bibitem{Rosner}
  J.~L.~Rosner,
  Phys.\ Rev.\  D {\bf 60} (1999) 074029
  [arXiv:hep-ph/9903543].

\bibitem{Cleo-iso}
  S.~Ahmed {\it et al.}  [CLEO Collaboration],
  Phys.\ Rev.\  D {\bf 66} (2002) 031101
  [arXiv:hep-ex/0206030].

\bibitem{BaBar-iso}B.~Aubert {\it et al.}\  [BaBar Collaboration],
  Phys.\ Rev.\  D {\bf 75}, 031101 (2007)
  [arXiv:hep-ex/0610027].

\bibitem{Ftopi-Ball}
  P.~Ball and R.~Zwicky,
  Phys.\ Rev.\  D {\bf 71} (2005) 014015
  [arXiv:hep-ph/0406232].

\bibitem{BtoD0KBelle}
  S.~K.~Swain {\it et al.}  [Belle Collaboration],
  Phys.\ Rev.\  D {\bf 68}, 051101 (2003)
  [arXiv:hep-ex/0304032].


\bibitem{BspsiphiCDF}
  CDF Collaboration, 
  CDF Note 10206.

\bibitem{BspsiphiD0}
  D0 Collaboration, 
  D0 Note 6098-CONF.

\bibitem{BsKK}R.~Fleischer and R.~Knegjens,
  arXiv:1011.1096 [hep-ph]; to appear in Eur.\ Phys.\ J. C.;
  R.~Fleischer,
  Eur.\ Phys.\ J.\  C {\bf 52}, 267 (2007)
  [arXiv:0705.1121 [hep-ph]].

\bibitem{BFK}
  K.~De Bruyn, R.~Fleischer and P.~Koppenburg,
  Eur.\ Phys.\ J.\  C {\bf 70}, 1025 (2010)
  [arXiv:1010.0089 [hep-ph]].

\bibitem{BspsiKS}CDF Collaboration,
CDF Note 10240; O.~Norniella  [CDF Collaboration],
  arXiv:1011.3531 [hep-ex].

\bibitem{chir}E.~E.~Jenkins and M.~J.~Savage,
  Phys.\ Lett.\  B {\bf 281}, 331 (1992).

\bibitem{BCN}P.~Blasi, P.~Colangelo, G.~Nardulli and N.~Paver,
  Phys.\ Rev.\  D {\bf 49}, 238 (1994)
  [arXiv:hep-ph/9307290].

\bibitem{BtoDpiCDF}
  A.~Abulencia {\it et al.}  [CDF Collaboration],
  Phys.\ Rev.\ Lett.\  {\bf 98} (2007) 061802
  [arXiv:hep-ex/0610045].

\bibitem{bib:CDF_fdfs3}CDF Collaboration, T. Aaltonen {\it et al.},
  Phys.\ Rev.\  D {\bf 77}, 072003 (2008)
  [arXiv:0801.4375 [hep-ex]].

\bibitem{bib:CDF_fdfs1}CDF Collaboration, F. Abe {\it et al.},
  Phys.\ Rev.\  D {\bf 60}, 092005 (1999).

\bibitem{bib:CKM2010}V. Gligorov, talk at CKM 2010, 6th International Workshop on the CKM 
Unitarity Triangle, Warwick, United Kingdom, September 6th--10th, 2010. 

%
%
%
\end{thebibliography}
\end{document}